\documentclass[twocolumn,aps,floatfix,showpacs,prl]{revtex4}
\usepackage{amsmath}
\usepackage{graphicx}
\usepackage{psfrag}
\usepackage[T1]{fontenc}

\begin{document}

\title{Exact diagonalization of the one dimensional Bose-Hubbard model \\ with local 3-body interactions}
\author{Tomasz Sowi\'nski}
\affiliation{
\mbox{Institute of Physics of the Polish Academy of Sciences, Aleja Lotnik\'ow 32/46, 02-668 Warsaw, Poland}}
\date{\today}
\begin{abstract}
In this Brief Report the extended Bose-Hubbard model with local two- and three-body interactions is studied by the exact diagonalization
approach. The shapes of the first two insulating lobes are discussed and the values of the critical tunneling for which the insulating phase loses stability 
for repulsive and attractive three-body interactions are predicted.  
\end{abstract}
\maketitle
Experimental progress on trapping and manipulating ultracold atoms confined in optical lattices has opened new perspectives for controlling many-body states of different quantum systems \cite{Zwerger,Lewenstein}. In the simplest case, when atoms interact mainly via two-body contact interactions, such systems are described in the context of the Bose-Hubbard (BH) model characterized by one free parameter only--the ratio of tunneling amplitude $J$ to the on-site two-body contact interaction strength $U$.  For such systems the quantum phase transition from superfluid (SF) to Mott insulator (MI) has been predicted and observed \cite{Zoller1,Greiner}. Depending on the physical realization there are many different extensions of standard BH models predicting the existence of novel quantum phases \cite{Lahaye,Mer,Dutta,Sowinski}.  One of the possible extensions of the BH model originates in taking into account not only two-body but also higher many-body interactions.  It is highly nonintuitive how many-body terms can appear in a description of systems in which all mutual interactions are purely two-body interactions. One should note however that direct interactions between particles are introduced as an effective description in the limit of low energy. Therefore, it is theoretically possible that in some particular experimental scenarios higher many-body terms can significantly change the properties of the system. One such scenario was proposed for polar molecules in optical lattices. As it is shown in  \cite{Buchler,Lauchli}, it is possible to control two- and three-body interactions between molecules independently. Another source of many-body terms in the presence of an optical lattice potential comes from the fact that mutual interactions can excite atoms to higher orbital states and modify the shape of the single-particle wave functions. This phenomenon can be effectively taken into account in the model by adding three-body interaction terms. In the case of repulsive forces, the single-atom wave functions are essentially broadened and therefore effective three-body terms become attractive \cite{Johnson,Bloch,Mark}. 

In this Brief Report, the ground state of the one-dimensional (1D) extended Bose-Hubbard model describing atoms interacting locally via two- and three-body interactions is studied. Such a model was previously discussed in the mean-field regime \cite{Chen, Zhou} and in the two-dimensional case \cite{Safavi}. Additionally, the first lobe of the insulating phase, i.e. when the average number of particles per lattice site $\rho=1$, was recently studied for the 1D case with density-matrix renormalization group (DMRG) methods \cite{Souza}. (The 1D model without three-body terms was studied in detail in \cite{Kuhner,Ejima2}.) Here we investigate  a physically more interesting region of the phase diagram when three-body interactions directly influence the properties of the ground state, i.e., when $\rho=2$. It is clear that in such a case, in contrast to the $\rho=1$ case, any tunneling process which leads to destruction of the insulating phase has to compete not only with two-body interactions but also with three-body ones. Moreover, in all previous analysis only repulsive three-body terms were taken into account. In the light of recent experiments \cite{Bloch} it is clear that the attractive case is probably more realistic and therefore I study also this case here. 

From now on I assume that higher orbital interactions can be effectively taken into account by adding a three-body interaction term. Therefore, I restrict myself to the lowest band of the optical lattice and then the system of $N$ bosons confined in a chain of $L$ lattice sites with periodic boundary conditions is described by the following extended Bose-Hubbard Hamiltonian
\begin{align}
H &= -J \sum_{i=1}^L  (\hat{a}_i^\dagger \hat{a}_{i+1}+\mathrm{H.c.}) + \frac{U}{2}\sum_{i=1}^L \hat{n}_i(\hat{n}_i-1) \nonumber \\
&+ \frac{W}{6}\sum_{i=1}^L \hat{n}_i(\hat{n}_i-1)(\hat{n}_i-2). \label{Hamiltonian}
\end{align}
The bosonic operator $\hat{a}_i$ ($\hat{a}^\dagger_i$) annihilates (creates) a particle at site $i$, $J$ is the 
nearest-neighbor hopping amplitude, and $U$ is the on-site two-body repulsion energy between bosons. The additional term
proportional to $W$ describes the on-site three-body interaction. For convenience, I introduce particle
number operators $\hat{n}_i=\hat{a}_i^\dagger\hat{a}_i$ and average occupation parameter $\rho=N/L$ where $N$ is the total number
of particles in the system. Note that the additional three-body term can be viewed as the first
occupation-dependent correction to the on-site two-body interaction in the standard Bose-Hubbard Hamiltonian
\begin{align}
\tilde{U}(n) &= U + \frac{W}{3}(n-2) + \cdots
\end{align}
Occupation dependence comes from the fact that for higher filling of the lattice site the lowest Wannier function does not properly describe the localized orbital. Similarly, the ground-state wave function of a harmonic oscillator does not resemble the ground-state solution of the Gross-Pitaevskii (GP) equation for large condensate fraction in the harmonic trap. Then, instead of expanding the ground-state solution in the basis of the trap potential, one can find the effective occupation-dependent frequency for which the harmonic oscillator ground-state mimics the GP solution. In the case of BH systems one can still use the lowest Wannier state but with occupation-dependent interactions and with appropriate renormalized interaction potential \cite{Johnson,Bloch,Mark}.

\begin{figure}
\includegraphics{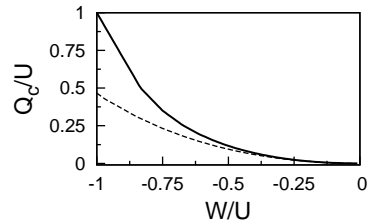}
\caption{The critical value of the amplitude of repulsive four-body interactions, $Q_c$, for which the system becomes stable when attractive three-body interactions are present in the system. The exact value of $Q_c$ depends on the filling (dashed line for $\rho=1$ and solid line for $\rho=2$) and it is obtained from the condition that the energy of $n>3$ particles occupying a chosen lattice site must be larger than the energy of the three-particle state at that site.} \label{Fig1}
\end{figure}

The Hamiltonian \eqref{Hamiltonian} is studied for repulsive three-body interactions ($W>0$) as well as for attractive ones ($W<0$). In the second case, it is necessary to take into account also four-body repulsive interactions,
\begin{equation}
\delta H = \frac{Q}{24}\sum_{i=1}^L \hat{n}_i(\hat{n}_i-1)(\hat{n}_i-2)(\hat{n}_i-3),
\end{equation}
to prevent the system collapsing. The value of the four-body interaction amplitude $Q$ does not affect the positions of the critical points, provided that it is large enough to prevent collapse. The critical value $Q_c$ of the four-body amplitude for which the studied system is stable can be easily determined. A sufficient condition is that the energy of any configuration with $n>3$ particles occupying a chosen lattice site must be larger than the energy of a configuration with three particles at that site. It is worth noticing that the value of $Q_c$ depends on an average filling $\rho$ since one has to compare the energies of whole lattice configurations instead of comparing on-site energy only. Its value for $\rho=1$ and $\rho=2$ are presented in Fig. \ref{Fig1}. In further analysis the value of $Q$ has been chosen very close to the critical value. In addition it was checked that the positions of critical points are insensitive to the precise value of $Q$. 

\begin{figure}
\includegraphics{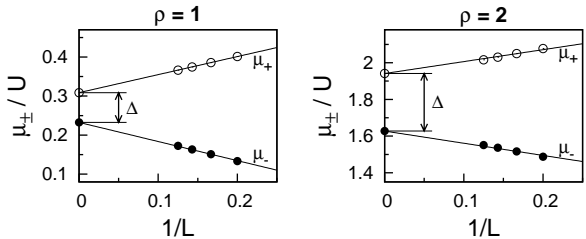}
\caption{The upper $\mu_+$ and lower $\mu_-$ limits of the insulating phase as a function of the inverse of the system size, $1/L$, for two densities $\rho=1$ (left) and $\rho=2$ (right) (Hamiltonian parameters are $W=U$ and $J/U=0.2$). The solid lines are linear fits to the numerical data points. Extrapolation to infinite system size $1/L \rightarrow 0$ gives the energy gap $\Delta$ of the insulating phase in the limit of infinite systems. } \label{Fig2}
\end{figure}

In this Brief Report, the diagram of quantum phases in the presence of the three-body interactions is studied theoretically. The numerical approach is based on the exact diagonalization of the Hamiltonian \eqref{Hamiltonian} and follows the approach discussed first by Elesin {\it et al.} \cite{Elesin}, adopted later to more sophisticated numerical techniques. The method draws on the observation that, for an infinite-size system, the MI phase, in contrast to the SF, has a nonzero energy gap $\Delta$ for adding or subtracting a particle to or from the system. Therefore, in principle one can determine insulating lobes in the phase diagram by finding ground-state energies of systems differing by one particle. It was shown by Fisher {\it et al.} \cite{Fisher} that for models considering on-site interactions only, the insulating phase can occur only for integer $\rho$. In such a case, the MI phase can be stable for nonzero hopping amplitude $J$. It can be easily shown with direct arguments that in the limit $J\rightarrow 0$ the energy gap of the first lobe ($\rho=1$) is not affected by three-body terms and it is equal to $\Delta = U$. In contrast, the second lobe ($\rho=2$) is highly influenced by three-body interactions and its width is equal to $\Delta = U+W$ for $J=0$. To find the energy gap $\Delta$ for an infinite system for given parameters $U$, $W$, and nonzero tunneling amplitude $J$, we follow a simple procedure. First, we perform exact diagonalization of the Hamiltonian for a finite size $L$ and a fixed number of particles, $N$. In this way we find the ground state of the system $|\mathtt{G}\rangle$, as well as the ground-state energy $E(L,N)$. Then we define the upper ($\mu_+$) and lower ($\mu_-$) limit of the insulating phase as follows
\begin{subequations}
\begin{align}
  \mu_+(\rho,L) &= E(L,\rho\,L+1)-E(L,\rho\,L), \\
  \mu_-(\rho,L) &= E(L,\rho\,L)-E(L,\rho\,L-1)
\end{align}
\end{subequations}
These quantities strongly depend on the size of the lattice. Since we are interested rather in their values in the limit of an infinite system, we diagonalize the Hamiltonian for different lattice sizes (in our calculations we take $L = 5,\ldots,8$). The diagonalization is performed without truncation in the full many-body basis; i.e. each lattice site can be occupied with $0,\ldots,L$ particles. Then we plot $\mu_+$ and $\mu_-$ as functions of $1/L$, and we extrapolate the data to the point $1/L = 0$. In this limit we define the energy gap of insulating phase $\Delta$ as a difference $\Delta=\mu_+-\mu_-$. To show how this method works in practice we present in Fig. \ref{Fig2} an example for $W=U$ and $J/U=0.2$ in two cases $\rho=1$ and $\rho=2$. As is seen, both $\mu$'s depend almost linearly on $1/L$ and therefore a linear data extrapolation gives quite good predictions. To test the accuracy of the obtained energy gap $\Delta$ we fit also higher-order polynomials in $1/L$ and compare the resulting values. The predicted values of $\Delta$ for linear, quadratic, and cubic polynomial fits are presented in Table \ref{Tabelka}. It is seen that the differences between the obtained values of the energy gap are about 2\% for $\rho=1$ and about $10\%$ for $\rho=2$. Therefore, for further calculations the validity of a linear dependence of the limiting $\mu$'s and $\Delta$ on $1/L$ is assumed.
\begin{table}
\begin{tabular}{c|cp{0.5cm}c}
\hline
Polynomial & \multicolumn{3}{|c}{$\Delta/U$} \\
order & $\rho = 1$ && $\rho=2$ \\
\hline
\hline
1 & $6.95 \times 10^{-2}$ && $2.55 \times 10^{-1}$\\
2 & $7.16 \times 10^{-2}$ && $2.85 \times 10^{-1}$\\
3 & $7.10 \times 10^{-2}$ && $2.75 \times 10^{-1}$\\
\hline 
\end{tabular}
\caption{The energy gap of the insulating phase in the limit of an infinite system size obtained for different fitting procedures. These values are calculated for the specific parameters of the Hamiltonian ($W=U$ and $J/U=0.2$) discussed as an example. } \label{Tabelka}
\end{table}

\begin{figure}
\includegraphics{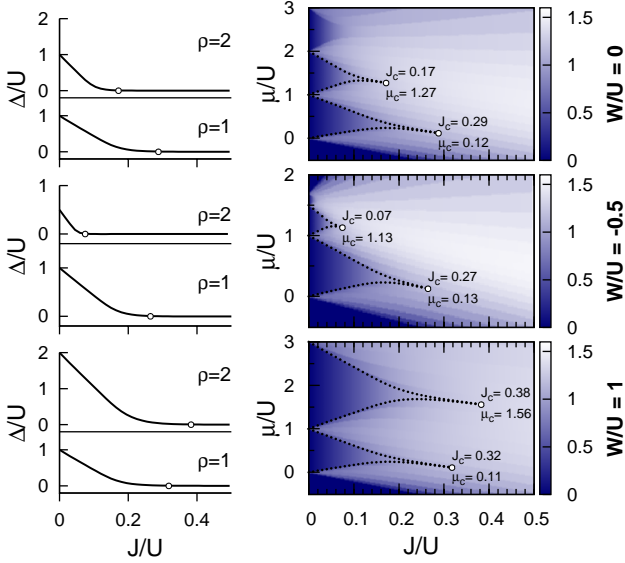}
\caption{(Color online) Properties of the system for different values of the three-body interaction strengths: $W=0$ (top), $W/U=-0.5$ (middle) and $W/U=1$ (bottom). On the left is the energy gap $\Delta$ of the first ($\rho=1$) and the second ($\rho=2$) MI lobe as a function of hopping amplitude $J$. On the right the phase diagrams of systems considered are presented. Dotted lines represent values of $\mu_-$ and $\mu_+$ for an infinite system obtained by extrapolation from finite system properties. Open circles are markers of the transition points, defined as the points where $\Delta$ is close to zero at the level of significance.  In the background of each phase diagram the density plots of the correlation function $\Phi$ are visualized. As was expected for $\rho=1$, the insulating phase does not change significantly, but for $\rho=2$ the size of the MI phase crucially depends on the three-body interaction parameter.} \label{Fig3}
\end{figure}

With the procedure described above, one can find the insulating gap $\Delta$ for different parameters of the Hamiltonian and different densities (left panel of Fig. \ref{Fig3}). The transition point from the MI to the SF phase occurs when $\Delta$ becomes equal to zero. Technically it is very hard to adopt this definition directly since $\Delta$ has some numerical uncertainty. In this Brief Report it is assumed that this uncertainty comes mainly from the extrapolating procedure and it is equal to $5\times 10^{-3}U$. As a consequence, we estimate the position of the transition point as the hopping amplitude for which the insulating gap becomes smaller than this uncertainty. These points are marked with open circles in Fig. \ref{Fig3}. The phase diagrams in the right-hand were obtained by plotting $\mu_\pm$ as functions of tunneling amplitude $J$. Additionally, in the background a density plot of the correlation function $\Phi=\langle\mathtt{G}| a^\dagger_ia_{i+1}|\mathtt{G}\rangle/N$ is shown. The meaning of this quantity is explained below. 

\begin{figure}
\includegraphics{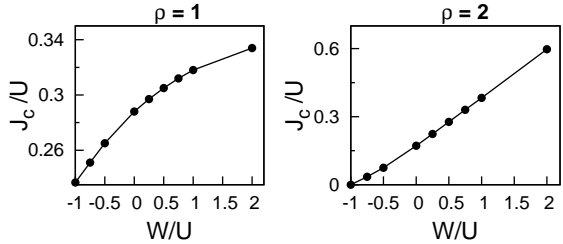}
\caption{Values of the critical tunneling as a function of the three-body interaction parameter $W$ for $\rho=1$ (left) and $\rho=2$ (right). In the first case, the position of the critical point in the phase diagram remains almost unchanged as was previously predicted \cite{Souza}. In the second case, the position of the critical point depends strongly on the three-body interactions. } \label{Fig4}
\end{figure}

For $W=0$ we recover the well-known phase diagram for the standard Bose-Hubbard model (top of Fig. \ref{Fig3}). The transition points for the first and second MI lobes occur for $J_c\approx 0.29U$ and $J_c\approx 0.17U$, respectively. These values are in good agreement with previous calculations \cite{Batrouni,Ejima}. For positive (negative) $W$ we observe that the transition point for the first MI lobe moves slightly towards larger (smaller) hopping amplitudes $J$ and smaller (larger) chemical potential $\mu$. Additionally, the energy gap $\Delta$ in the limit of vanishing tunneling is equal to $U$ for any $W$. This is quite obvious since in this limit every lattice site is occupied by exactly one boson and therefore three-body interactions can be completely neglected. In this way we confirm observations based on DMRG methods \cite{Souza}. The situation is quite different for the second insulating lobe. In this case, for repulsive three-body interactions ($W>0$), the SF phase occurs for larger tunneling amplitudes. It comes from the fact that tunneling, which destroys the insulating phase, has to overcome an energy barrier increased by $W$.  In the opposite case, for attractive three-body interactions ($W<0$) the destruction of the MI phase should be much easier, since then the energy of a three-body configuration is decreased by $W$. It is very simple to show that in the limit of vanishing $J$ the energy gap $\Delta$ (the width of the MI lobe in the phase diagram in the $\mu$ direction) should be equal to $U+W$. This means that in the limit $W\rightarrow -U$ a vanishing of the second insulating lobe should be observed. Numerical results fully confirm all these predictions (Fig. \ref{Fig3}). In Fig. \ref{Fig4} we show the dependence of critical hopping amplitude on the strength of the three-body interactions. As was suspected, the critical tunneling $J_c$ is much more sensitive to three-body interactions for the second lobe than for the first one.

\begin{figure}
\includegraphics{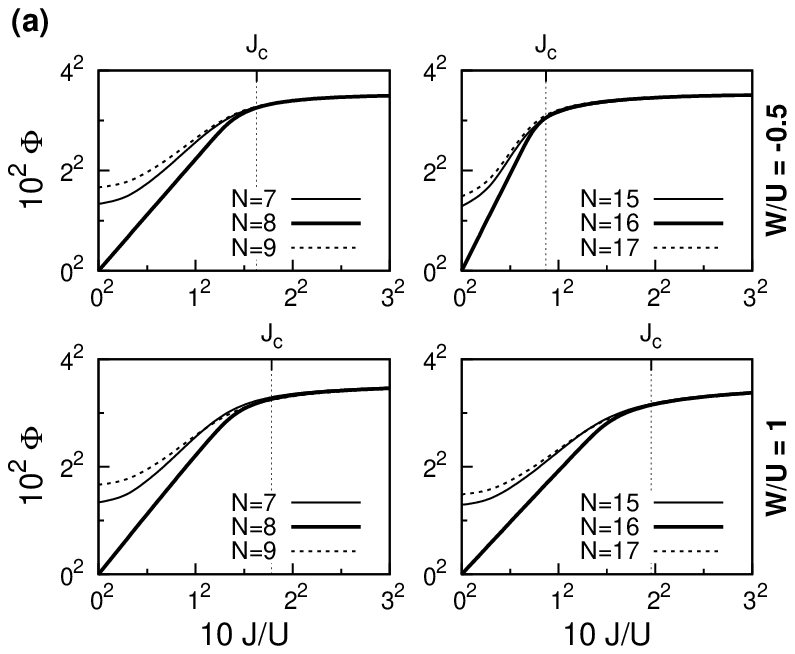}
\includegraphics{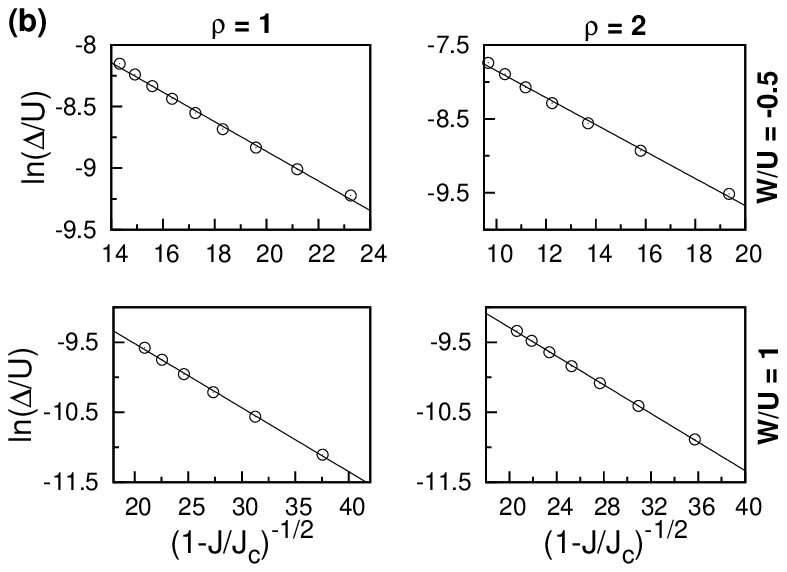}
\caption{(a) The correlation $\Phi=\langle\mathtt{G}| a_{i+1}^\dagger a_i|\mathtt{G}\rangle/N$ as a function of tunneling amplitude $J$ for different Hamiltonian parameters and fillings (system size $L=8$). It is seen that for integer fillings (bold solid lines), in contrast to other noninteger ones (dashed and thick solid lines), the function $\Phi$ decays to zero in the limit $J\rightarrow 0$. The transition tunneling $J_c$ predicted by the other method occurs when the $\Phi$ curve for integer filling detaches from the curves for fractional fillings. This behavior is universal for repulsive ($W>0$) as well as for attractive ($W<0$) three-body interactions. Note that for legibility a nonlinear scaling has been used. (b) The decay of the energy gap $\Delta$ in the neighborhood of the critical point for repulsive (top) and attractive (bottom) three-body interactions. In chosen variables the data points fit the linear predictions of the Kosterlitz-Thouless phase transition. } \label{Fig5}
\end{figure}

The exact diagonalization method gives quite a nice confirmation that the MI phase can occur only for integer fillings. One can calculate the correlation function $\Phi=\langle\mathtt{G}| a_{i+1}^\dagger a_i|\mathtt{G}\rangle/N$ which describes hopping of particles between neighboring sites, and it might be viewed as a number-conserving analog of the mean field. Due to translational invariance this quantity does not depend on the index $i$. From numerical calculations for finite lattices we find that for vanishing tunneling amplitudes $J$ this quantity drops exactly to zero only for integer fillings. In other cases it remains nonzero even in the limit $J\rightarrow 0$. Moreover, the critical tunneling $J_c$ determined previously from the energy-gap arguments occurs when $\Phi$ (for integer filling) starts to decay much faster than for surrounding noninteger fillings. In Fig. \ref{Fig5}(a), values of $\Phi$ for different fillings and three-body interactions are presented (the number of lattice sites is $L=8$). Additionally, in Fig. \ref{Fig3} the density plot of $\Phi$ is visualized in the background of appropriate phase diagrams. The darker areas correspond to smaller values of $\Phi$. In these cases, $\Phi$ is calculated by exact diagonalization of the grand-canonical Hamiltonian $H-\mu N$ in a truncated basis with $L=8$. The insulating lobes predicted this way match the lobes determined by energy-gap boundaries.

Finally, let me note that the standard one-dimensional Bose-Hubbard model has totally different properties than those predicted by the mean-field approximation \cite{Giamarchi}. This originates from the fact that at the tip of the Mott lobe the one-dimensional BH model belongs to the universality class of the $XY$ spin model. This means that the phase transition from the MI to the SF phase is of the Kosterlitz-Thouless (KT) type. It is characterized by the exponential decay of the correlation length \cite{Kuchner}. In other words, the energy gap $\Delta$ decays exponentially in the neighborhood of the critical point; i.e., $\ln(\Delta/U)$ is a linear function of $1/\sqrt{1-J/J_c}$ for $J<J_c$. Numerical results based on an exact diagonalization method show that this behavior does not depend on the three-body interactions and the KT transition occurs in the system also in the case when three-body interactions (attractive as well as repulsive) are taken into account. Example results for $W=-0.5U$ and $W=U$ are presented in Fig. \ref{Fig5}(b). The numerical predictions fit almost perfectly to the theoretical predictions of KT phase transition. 

To conclude, in this Brief Report the extended Bose-Hubbard model with additional local three-body interactions was studied. To find the critical behavior of the system, exact diagonalization of the Hamiltonian was used. In this way, the phase diagrams for repulsive and attractive three-body interactions were obtained. It was shown that the shape of the second insulating lobe (for two bosons per lattice site on average), in contrast to the first one, depends  crucially on the three-body interactions. Additionally, it was checked that local three-body interactions do not change the characteristic properties of the phase transition. 

The author thanks M. Gajda and P. Deuar for fruitful comments and suggestions. This research was funded by the National Science Center from Grant No. DEC- 2011/01/D/ST2/02019  and the EU STREP NAME-QUAM.

\end{document}